\documentclass[12pt]{iopart}
\usepackage{epsfig}
\usepackage{rotating}

\begin{document}

\title[Exchange interaction in coupled quantum dots]
{Effect of confinement potential shape on exchange interaction in coupled quantum dots}
\author{A Kwa\'sniowski and J Adamowski}
\address{Faculty of Physics and Applied Computer Science, AGH University of Science and Technology,
al. Mickiewicza 30, 30-059 Krak\'ow, Poland}

\date{\today}

\ead{kwasniowski@novell.ftj.agh.edu.pl}

\begin{abstract}
Exchange interaction has been studied for electrons
in coupled quantum dots (QD's) by a configuration interaction method
using confinement potentials with different profiles.
The confinement potential has been parametrized by
a two-centre power-exponential function,
which allows us to investigate various types of QD's described by
either soft or hard potentials of different range.
For the soft (Gaussian) confinement potential the exchange energy
decreases with increasing interdot distance due to
the decreasing interdot tunnelling.
For the hard (rectangular-like) confinement potential we
have found a non-monotonic behaviour
of the exchange interaction as a function of distance
between the confinement potential centres.
In this case, the exchange interaction energy exhibits a pronounced maximum
for the confinement potential profile which corresponds to
the nanostructure composed of the small inner QD
with a deep potential well
embedded in the large outer QD with a shallow potential well.
This effect results from the strong localization of electrons
in the inner QD, which leads to the large singlet-triplet splitting.
Implications of this finding for quantum logic operations
have been discussed.
\end{abstract}


\maketitle

\section{Introduction}

An exchange interaction between electrons localized in coupled
quantum dots (QD's) is a very promising tool for
manipulating qubits in semiconductor nanodevices \cite{loss98}.
This interaction can change the spin of the electron, which
allows us to perform the quantum logic operations with spin
qubits \cite{loss98,burk99,burk00,hu01,bell04}.
Recently, the exchange-interaction induced spin swap operations
in coupled QD's have been simulated by a direct solution of a time-dependent
Sch\H{o}dinger equation \cite{mosk07}.
The quantum logic operations can be performed
in the laterally coupled QD's
\cite{elz03,petta,wang06} and quantum wire QD systems \cite{fuhrer07,zh07}.
Conditions of a realization of the logic operations with qubits in QD's
are determined by the properties of electron states in QD nanostructures.
The electron states of the laterally coupled QD's
have been studied theoretically in Refs.
\cite{yan01,ront04,har02,reim02,bs04,zh06,mar03,wen00,stopa,ped07,naga99}.
In electrostatically gated QD's \cite{elz03},
the properties of the electron quantum states
can be tuned by changing the external voltages applied to the gates \cite{hand06}.

In order to perform the high-fidelity quantum logic operations
with spin qubits the exchange interaction should be possibly strong.
The exchange interaction energy, defined as the difference between the
lowest triplet and singlet energy levels,
depends on the localization of electrons in the QD's.
In general, the stronger the electron localization
the stronger the exchange coupling.
The electron localization is determined by the profile of the
potential confining the electrons within the QD's.
Therefore, the exchange interaction depends
on the shape and range of the confinement potential.
Usually, the confinement potential in coupled QD's is modelled
by the two-centre parabolic \cite{har02,ped07} or Gaussian \cite{bs04,zh06} potential.

In the present paper, we propose the two-centre
power-exponential (PE) potential \cite{ciurla}, which allows us
to study a broad class of confinement potentials with
different shapes. The one-centre PE potential \cite{ciurla}
is very well suited for a description of the electrostatic QD's \cite{hand06,lis03}.
It has a form
\begin{equation}
V(r) = -V_0 \exp [-(r/R)^p] \; ,
\label{PE}
\end{equation}
where $V_0$ is the depth of the potential well, $R$ is the range of the potential,
which determines the QD size, and parameter $p \geq 2$ describes
''a softness'' of the potential, i.e., a smoothness of the QD boundaries.
If $p \simeq 2$, the potential is ''soft'', and if $p \geq 4$, the potential
is ''hard'', i.e., it possesses the walls with large steepness.
Parameter $p$ can be used to describe a different steepness of the
QD boundaries.  Therefore, PE potential (\ref{PE})
can be applied to a modelling of the electrostatically
gated QD's \cite{elz03,hand06} and self-assembled QD's with compositional
modulation \cite{siv98}.
The influence of the smoothed interfaces on the electronic and optical properties of
GaAs/$\mathrm{Al_x Ga_{1-x} As}$ QD's
have been studied in a recent paper \cite{mlin07}.

The exchange interaction can be controlled by internal nanostructure parameters,
e.g., size and geometry of the coupled QD's, and
external electric and magnetic fields \cite{bs04,zh06,stopa}.
It has been shown \cite{bs04} that the asymmetry
of QD's leads to the increase of this interaction.
Recently, the size effects in the exchange coupling
have been studied for quantum wire QD systems \cite{zh07}.

In the present paper, we investigate the influence of the shape and range
of the confinement potential on the exchange
interaction in laterally coupled QD's using the two-centre
PE potential.
The paper is organized as follows: a theoretical model is presented in
Section II, the calculation methods and results are given in Section III,
Section IV includes the discussion and
Section V -- the conclusions and summary.

\section{Theory}

We study the system of two electrons confined in laterally coupled QD's
with identical confinement potentials.
The geometry of the nanostructure, which consists of two separated QD's,
is illustrated in Figure 1.
We describe each QD by the two-dimensional (2D) rotationally symmetric potential
well centred at positions $\pm\mathbf{a}$, where $\mathbf{a}=(d/2,0)$
and $d$ is the distance between the centres of the potential wells.

\begin{figure}[htbp]
\begin{center}
\epsfig{figure=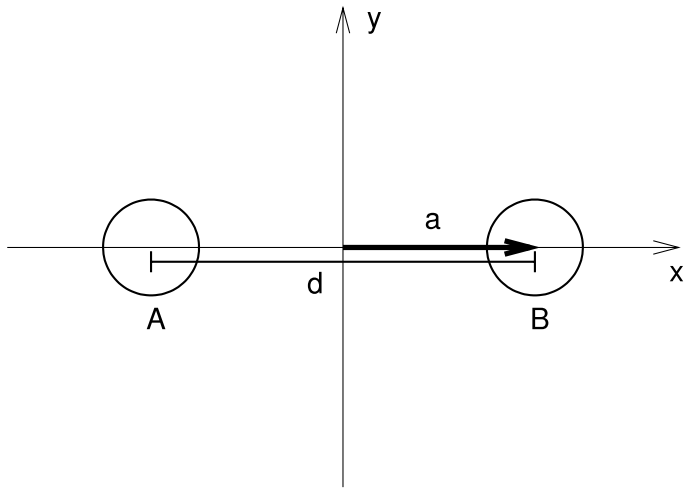,width=90mm}
\end{center}
\caption{Schematic of the coupled QD system.  Circles $A$ and $B$ represent
the QD's,  $d$ is the distance between the centres of the confinement
potentials, $\mathbf{a}$ is the position vector of the centre of the right QD.}
\end{figure}

In the effective mass approximation the Hamiltonian of the system reads
\begin{equation}
H = h_1 + h_2 + \frac{e^2}{4\pi\varepsilon_0\varepsilon_s r_{12}} \;,
\label{H_tot}
\end{equation}
where $h_j$ ($j=1,2$) is the one-electron Hamiltonian,
$\varepsilon_0$ is the electric permittivity of the vacuum,
$\varepsilon_s$ is the static relative electric permittivity,
$r_{12}= |\mathbf{r}_1-\mathbf{r}_2|$ is the electron-electron distance,
and $\mathbf{r}_j=(x_j,y_j)$ denote the electron position vectors.
The one-electron Hamiltonian has the form
\begin{equation}
h_j = -\frac{\hbar^2}{2m_e}\nabla_j^2 + V(\mathbf{r}_j) \;,
\label{h1}
\end{equation}
where $m_e$ is the electron effective band mass and $V(\mathbf{r}_j)$ is the confinement potential.
We assume that the effective mass and the static electric
permittivity do not change across the QD boundaries.
This assumption is well justified
for the electrostatic QD's based on GaAs \cite{hand06}.
For the coupled QD's the confinement potential is taken on as a sum
of PE potentials [Eq.~(\ref{PE})] centred at $\pm \mathbf{a}$.
In the explicit form,
\begin{equation}
V(\mathbf{r}) = -V_0\Big\{\exp[-(r_A/R)^p] + \exp[-(r_B/R)^p]\Big\} \;,
\label{conf}
\end{equation}
where $r_A = |\mathbf{r}+\mathbf{a}|$ and $r_B = |\mathbf{r}- \mathbf{a}|$.

Formula (\ref{conf}) defines a broad class of two-centre confinement potentials
with different shapes.
For fixed potential well depth $V_0$ and range $R$ parameter
$p$ characterizes the softness (hardness) of the confinement potential.
The smaller (larger) $p$ the potential is more soft (hard).
For $p=2$ the confinement potential has the Gaussian shape.
This potential is soft, i.e., the potential walls at
the QD boundaries, have fairly small steepness and are partly penetrable for the electron.
For $p \geq 4$ the confinement potential becomes hard,
i.e., the potential walls at the QD boundaries are steep.
For $p \geq 10$ we deal with the rectangular-type, very hard, confinement potential.
Figure 2 shows the different profiles of
the two-centre confinement potential [Eq. (\ref{conf})].
In Figure 2 and throughout the present paper we are using the donor Rydberg
$R_D=m_e e^4/(32\pi^2 \varepsilon_0^2\varepsilon_s^2\hbar^2)$
as the unit of energy and the donor Bohr radius $a_D=4\pi\varepsilon_0\hbar^2/(m_e e^2)$
as the unit od length.
For GaAs, $R_D \simeq 6$ meV and $a_D \simeq 10$ nm.
We note the essential difference in potential profiles
between the soft ($p=2$) and hard ($p=10$) confinement potentials
for intermediate distances $d$, i.e., for $d \simeq 3 \; a_D$ in Figure 2.
In the case of soft confinement potential,
the resulting two-centre potential changes
smoothly with increasing $d$ from the single to the double potential well.
In this case,
the increasing intercentre distance $d$ leads
first to the increase of the potential well size
and next to the formation of the potential barrier
for $d \geq 3 \; a_D$.
If the confinement potential is hard,
then -- for intermediate intercentre distances $d$ --
we obtain a narrow deep potential well surrounded
by fairly wide potential steps, on which the potential
is flat [cf. Figures 2(b,c)].
This shape of the confinement potential
corresponds to the core-shell QD-nanostructure, which was
realized in CdTe/CdS self-assembled QD's \cite{core_shell}.
In this case, we deal with the compound QD nanostructure,
which consists of the small inner QD embedded in the
larger outer QD.

\begin{figure}[htbp]
\begin{center}
\begin{turn}{-90}
\epsfig{figure=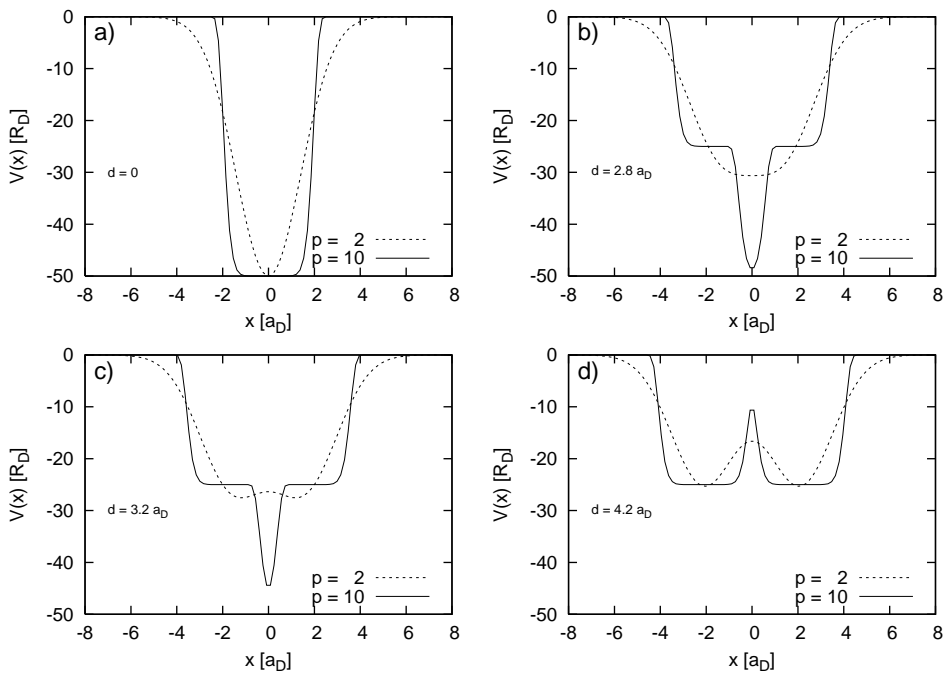,width=90mm}
\end{turn}
\end{center}
\caption{Confinement potential profile for different distances $d$
between the potential well centres as a function of coordinate $x$.
Solid curves correspond to the hard confinement potential ($p=10$),
dashed curves correspond to the soft confinement potential ($p=2$).
The plots for $V_0=25 R_D$ and $R=2 a_D$.  }
\end{figure}

\section{Results}

We are mainly interested in the influence of the shape of the confinement potential,
in particular its softness, on the electronic properties of coupled QD's.
Therefore, we have performed the calculations for fixed depth $V_0 = 25 \; R_D$ and range
$R = 2 \; a_D$ (except otherwise specified).
These values of the parameters correspond to laterally coupled GaAs QD's \cite{wang06}.

\subsection{One-electron problem}

The one-electron eigenvalue problem for the one-centre
confinement potential has been solved for the Gaussian potential
in Ref. \cite{gauss} and  for the PE potential in Ref. \cite{ciurla}.
In the present paper, we consider the one-electron bound states
for the two-centre confinement potential (\ref{conf}).
We have solved the eigenvalue problem for Hamiltonian
(\ref{h1}) by the imaginary time step method \cite{ITS}
applying the finite-difference approximation of Hamiltonian (\ref{h1})
on the 2D grid with $101\times 101$ mesh.  In this case,
the accuracy of the method \cite{ITS}
is high enough to treat the obtained numerical solutions as exact.

Figure 3 shows the results for the six lowest-energy levels,
which correspond to the one-electron states used to
a construction of two-electron configurations.
Figures 3(a) and 3(b) display the results for the soft ($p=2$) and hard ($p=10$) confinement potential,
respectively.  For $d=0$ we deal with the single QD with the potential well
of double depth ($2V_0$), while for large $d$ the QD's are separated by
the potential barrier.  For intermediate $d$ the superposition of hard-wall potentials
corresponds to the inner-outer QD nanostructure [cf. Figures 2(b) and 2(c)].
These properties of the confinement potentials affect the one-electron
states (Figure 3).
The one-electron energy levels are monotonically increasing functions of
interdot distance $d$.
For $d=0$ the excited-state energy levels exhibit
a degeneracy, which is removed for $d > 0$.
We note that for intermediate values of $d$, i.e., for $d \simeq 3 \; a_D$,
the energy levels quickly increase with increasing $d$.
For large $d$ we obtain only two degenerate energy levels,
which correspond to the electron bound in the single QD.

\begin{figure}[htbp]
\begin{center}
\epsfig{figure=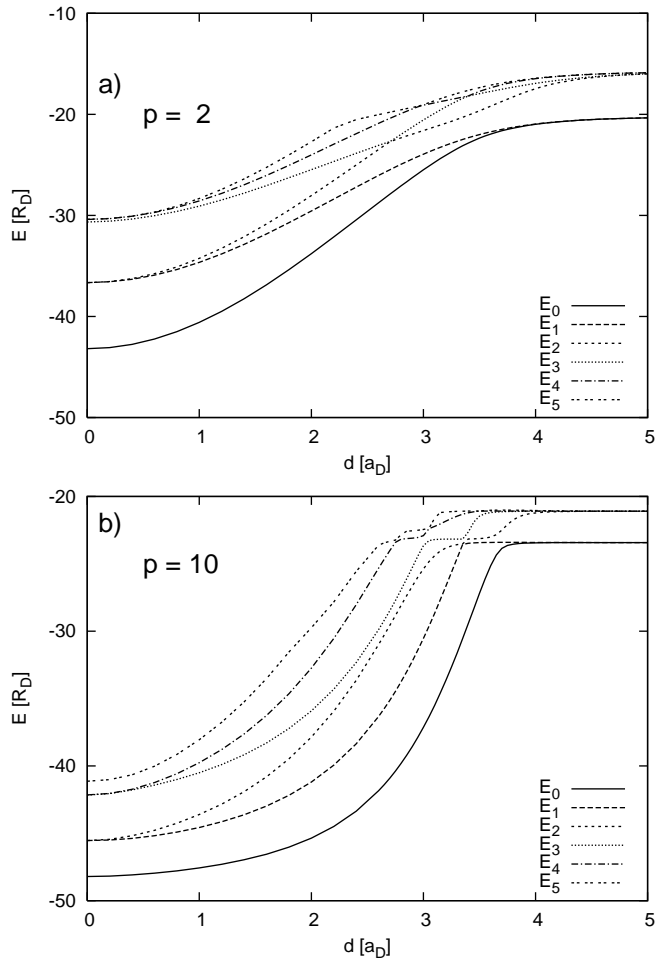,width=90mm}
\end{center}
\caption{Lowest-energy levels $E_0, \ldots, E_5$ of the single electron
confined in coupled QD's as functions of
distance $d$ between the confinement potential centres for (a) soft confinement potential with $p=2$
and (b) hard confinement potential with $p = 10$.}
\end{figure}

The dependence of the ground-state energy $E_0$
on the softness parameter $p$ is depicted in Figure 4.
If the confinement potential is more hard, i.e., for large $p$,
the ground-state energy takes on the lower values, which means that
the electron is more strongly bound due to the larger
quantum ''capacity'' of the QD.

\begin{figure}[htbp]
\begin{center}
\epsfig{figure=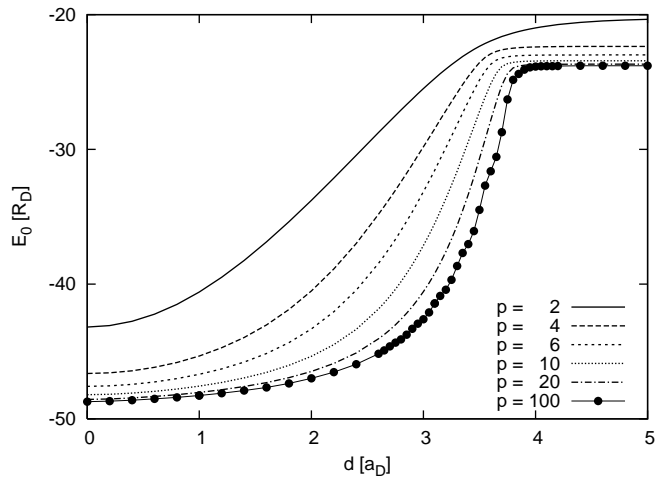,width=90mm}
\end{center}
\caption{Ground-state energy $E_0$ of the single electron
confined in the coupled QD's as a function of interdot distance $d$
for different softness $p$ of the confinement potential.}
\end{figure}

\subsection{Two-electron problem}

Two electrons confined in the coupled QD's form molecular-like states,
called the artificial molecules \cite{yan01,ront04}.
We solve the two-electron eigenvalue problem by the configuration
interaction (CI) method using the one-electron numerical solutions
obtained in the previous subsection.
Augmenting the calculated spatial wave functions by
the eigenfunctions of the $z$ component of the electron spin
we obtain the one-electron spin-orbitals $\psi_{\nu\sigma}(\mathbf{r})$,
where $\nu$ and $\sigma$ are the orbital and spin quantum numbers, respectively.
Next, we construct Slater determinants
\begin{equation}
\chi_n(\mathbf{r}_1,\mathbf{r}_2)
=\mathcal{A}[\psi_{\nu_1\sigma_1}(\mathbf{r}_1)\psi_{\nu_2\sigma_2}(\mathbf{r}_2)] \;,
\label{Slater}
\end{equation}
where $\mathcal{A}$ is the antisymmetrization operator and $n$ labels
different two-electron configurations with well-defined total spin.
According to the CI method the two-electron wave function is a linear combination of Slater
determinants (\ref{Slater})
\begin{equation}
\Psi(\mathbf{r}_1,\mathbf{r}_2) = \sum\limits_n c_n \chi_n(\mathbf{r}_1,\mathbf{r}_2) \;.
\label{CI}
\end{equation}
In the present calculations, we have used 15 Slater determinants,
which were constructed from the six lowest-energy one-electron states.
We have checked that including 20 Slater determinants
in expansion (\ref{CI}) only slightly improves the results,
but the computation time considerably increases.
Two-electron Hamiltonian (\ref{H_tot}) has been diagonalized in basis (\ref{CI}).
All the matrix elements, including the electron-electron interaction energy,
have been calculated numerically on the 2D grid defined for the one-electron problem.

\begin{figure}[htbp]
\begin{center}
\epsfig{figure=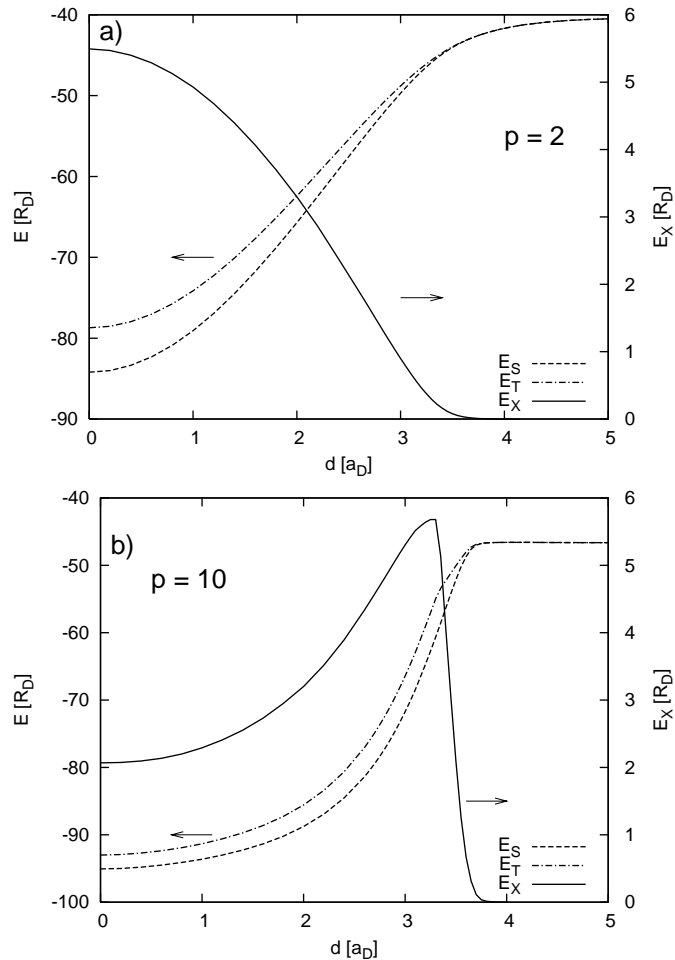,width=90mm}
\end{center}
\caption{Lowest-energy levels of the singlet ($E_S$) and triplet ($E_T$) states
(left scale) and exchange energy $E_X$ (right scale) as functions of distance $d$
between the centres of the confinement potential for
(a) soft ($p=2$) and (b) hard ($p=10$) confinement potential.}
\end{figure}

The results for the lowest-energy levels of the singlet ($E_S$) and triplet ($E_T$) states
are displayed in Figure 5, which also shows the exchange interaction energy $E_X$ energy
defined as
\begin{equation}
E_X = E_T-E_S \;.
\label{ex}
\end{equation}
The energy of the singlet and triplet states increases monotonically
with increasing distance $d$ between the confinement potential centres.
Therefore, the binding energy, defined as $E_B=-E_{T,S}$, of both the states
is a decreasing function of interdot separation,
i.e., the two-electron states are more weakly bound for
more separated QD's.  Figure 5 shows the different behaviour
of the the exchange energy for the soft and hard confinement potential.
If the confinement potential is soft [cf. Figure 5(a) for $p=2$],
the exchange energy decreases with increasing intercentre distance
and for large $d$ is negligibly small.
This behaviour results from the decreasing interdot tunnelling
with increasing distance between the QD centres.
If the confinement potential is hard [cf. Figure 5(b) for $p=10$],
the exchange energy increases with intercentre distance (for small $d$),
exhibits a pronounced maximum for $d\simeq 3.3 a_D$ (for $p=10$),
and next rapidly falls down to zero
for large $d$.  We shall discuss the origin of this behaviour
in the next subsection.

\begin{figure}[htbp]
\begin{center}
\epsfig{figure=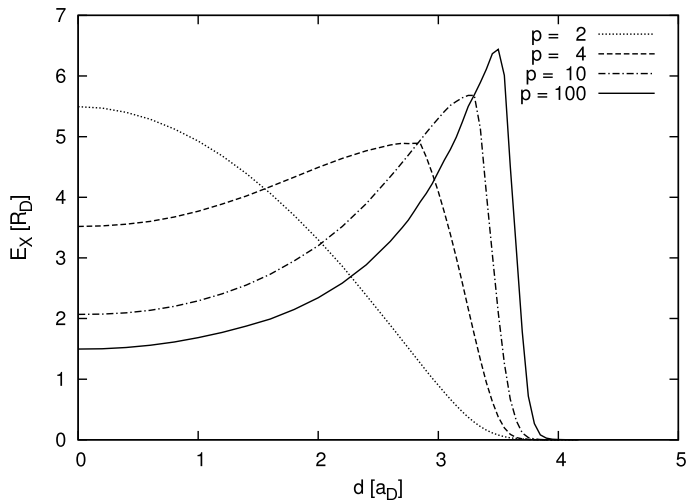,width=90mm}
\end{center}
\caption{Exchange energy $E_X$ as a function of distance $d$
between the confinement potential centres for different values of softness parameter $p$.}
\end{figure}

Figure 6 shows the dependence of the exchange energy on intercentre distance $d$ and softness
$p$ of the confinement potential.  We see that the maximum of the exchange energy
is more pronounced if parameter $p$ is large, i.e., the confinement potential is hard.
Moreover, the maximal value of the exchange energy
is larger if the confinement potential is more hard.

\subsection{Electron density distribution}

In order to explain the different behaviour of the
exchange energy for the soft and hard confinement potentials,
we have calculated the one-electron probability density
\begin{equation}
\rho_1(\mathbf{r})=\sum_{i=1}^2\int d^2 r_1 d^2 r_2
\Psi^\ast(\mathbf{r}_1,\mathbf{r}_2)
\delta(\mathbf{r}-\mathbf{r}_i) \Psi(\mathbf{r}_1,\mathbf{r}_2)
\label{ro1}
\end{equation}
and displayed it in Figures 7, 8, and 9.

\begin{figure}[htbp]
\begin{center}
\epsfig{figure=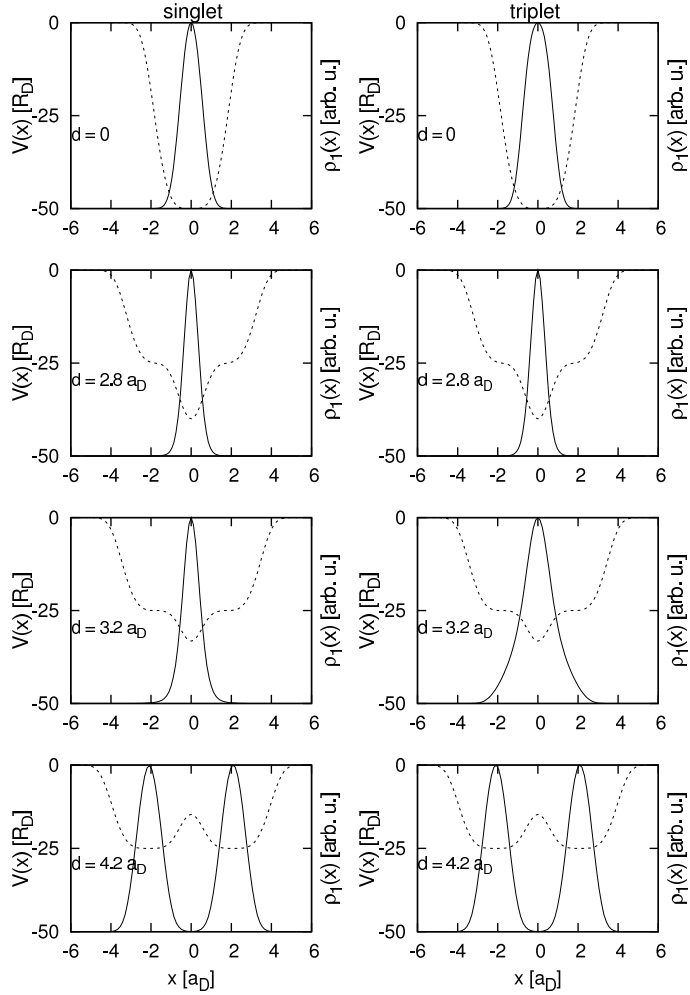,width=90mm}
\end{center}
\caption{One-electron probability density $\rho_1$ (solid curves) and confinement potential $V$
(dashed curves) for $p=4$ as functions of $x$ for $y=0$ for the singlet (left panel)
and triplet (right panel) states.}
\end{figure}

\begin{figure}[htbp]
\begin{center}
\epsfig{figure=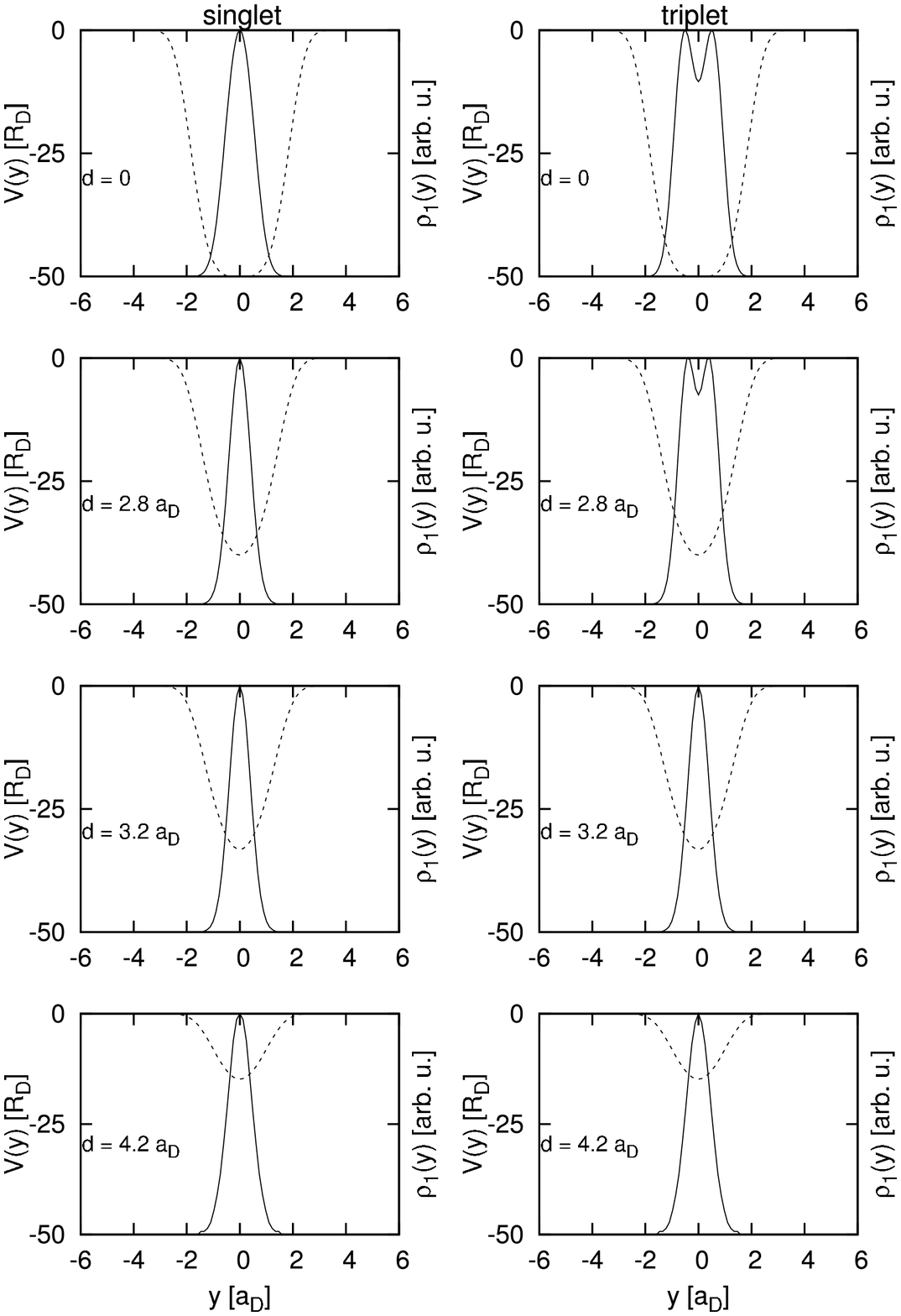,width=90mm}
\end{center}
\caption{One-electron probability density $\rho_1$ (solid curves) and confinement potential $V$
(dashed curves) for $p=4$ as functions of $y$ for $x=0$ for the singlet (left panel)
and triplet (right panel) states.}
\end{figure}

The properties of the electron density
are determined by the shape of the total confinement potential,
which is different in the $x$ and $y$ directions.
The $x$-dependence of the one-electron probability density
(Figure 7) is very similar for the singlet and triplet states.
In the $x$ direction, the electrons are localized in the single potential well
for small $d$ (including $d=0$).
For large interdot separation (cf. Figure 7 for $d=4.2 a_D$)
the electrons are localized with equal probabilities in the two quantum wells
separated by the energy barrier.
At the intermediate distances, i.e., for $d \simeq 3 a_D$,
the confinement potential profile corresponds to the core-shell QD nanostructure
with the inner and outer QD's (cf. Figure 7 for $d=2.8 a_D$ and $3.2 a_D$).
In this case, the electrons are mainly
localized within the inner QD with the deep potential well.
Figure 7 shows that -- for all interdot distances --
the electron localization in the singlet state is stronger
then in the triplet state.  In particular,
we observe that -- for $d=3.2 a_D$ -- the triplet electron density is fairly large
in the outer QD region.

Figure 8 displays the shape of the confinement potential
and the one-electron probability density as functions of $y$
for $x=0$.  For $d < 3 a_D$ we observe the essential qualitative difference
in the $y$-dependent electron distribution between the singlet
and triplet states.
In the singlet state, the electron distribution in the $y$ direction
is similar to that in the $x$ direction.
However, in the triplet state, the electron density distribution
is more spread out and exhibits two pronounced maxima separated
by a minimum at $y=0$.
For sufficiently large interdot distances ($d > 3 a_D$), the electron density
distributions for the singlet and triplet states are again
very similar.

The comparison of Figures 7 and 8 for $d < 3 a_D$ allows us to
extract the information about the distribution of the electrons in the triplet state.
The electrons with the same spin are strongly localized
in the inner QD in the $x$ direction (Figure 7), but are
more weakly localized in the $y$ direction (Figure 8).
The weaker localization of the electrons in the $y$ direction results from
the larger effective range of the potential well
in this direction (cf. dashed curves in Figures 7 and 8).
Figures 7 and 8 show the reasons of the existence
of the maximum exchange interaction for the hard confinement potential.
We see that the increase of the exchange coupling is caused
by the strong electron localization in the inner QD
and the considerable difference of the electron density distributions for the singlet
and triplet states.  These effects lead to the increase of the triplet-singlet energy difference,
which in turn results in the formation of the pronounced maximum of exchange energy
at intermediate distances between the confinement potential centres (cf. Figure 5).

\begin{figure}[htbp]
\begin{center}
\epsfig{figure=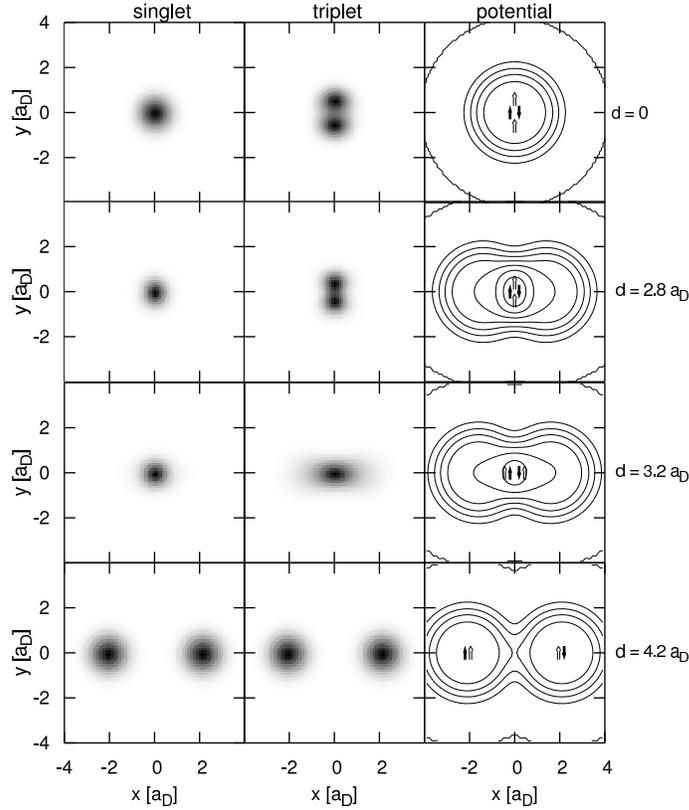,width=90mm}
\end{center}
\caption{Contours of the one-electron density distribution
for the singlet (left panel) and triplet (middle panel) states,
and the corresponding confinement potential profiles (right panel)
for different distances $d$ between the potential well centres
in the case of the hard confinement potential with $p=4$ and $R=2 a_D$.
Full (empty) arrows in the right panel show the sites with the maximum
electron density for the singlet (triplet) states.}
\end{figure}

These effects are additionally illustrated in Figure 9,
which displays the contour plots of the one electron density
distribution and confinement potential on the $x-y$ plane.
The arrows schematically show the sites with the largest electron density
for the singlet and triplet states.
The plots for $d=0$ and $d=2.8 a_D$ show
that -- in the triplet state -- each electron
is localized at a different site.
The plots for $d=2.8 a_D$ correspond to the maximum of the
exchange energy for $p=4$.
The triplet electron density distribution is anisotropic on the $x-y$ plane.
For small $d$ the electrons with the same spin are aligned in the $y$ direction.
This configuration rapidly changes for $d \simeq 3 a_D$ and
the triplet density distribution
becomes more spread out in the $x$ direction (cf. Figure 9 for $d=3.2 a_D$).
Figures 7, 8, and 9 show that
-- for large interdot distances --
the electron density distribution is the same in the singlet and triplet
which leads to the singlet-triplet degeneracy, i.e.,
the exchange energy tends to zero for large $d$.

Figure 10 displays the dependence of maximum exchange energy
$E_X^{max}$ (cf. Figure 6) on parameter $p$ and range $R$ of the confinement
potential.  The maximum exchange energy increases with
increasing parameter $p$, i.e., increasing hardness of the confinement
potential, and is the largest for the rectangular-like potential well.
However, the maximum triplet-singlet splitting quickly decreases with
increasing range $R$, i.e., increasing QD size.
The $R$-dependence of $E_X^{max}$ allows us
to determine the size effect in the exchange interaction \cite{zh07}.
This dependence can be parametrized as follows:
\[ E_X^{max}(R)=a/R + b \;, \]
with $a=9.975$ and $b=-0.073$ (in donor units).

\begin{figure}[htbp]
\begin{center}
\begin{turn}{-90}
\epsfig{figure=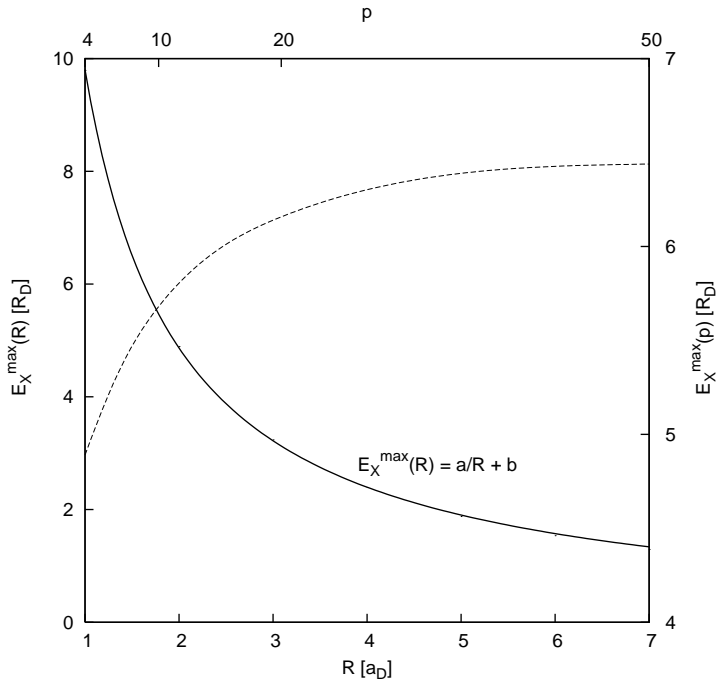,width=90mm}
\end{turn}
\end{center}
\caption{Maximum exchange energy $E_X^{max}$ as a function of
range $R$ of the confinement potential (solid curve for $p=4$)
and softness parameter $p$ (dashed curve for $R=2 a_D$)
calculated for $V_0 = 25 \; R_D$.}
\end{figure}

For fixed parameters $p$ and $V_0$ the maximum exchange energy depends
not only on $R$, but also on distance $d$ between the centres
of the confinement potential.
Therefore, the intercentre distance $d=d^{max}$,
which corresponds to the maximum of the exchange energy,
changes along the curves $E_X^{max}(R)$ and $E_X^{max}(p)$ in Figure 10.
The dependence of $d^{max}$ on parameters $R$ and $p$
is plotted in Figure 11.   It appears that intercentre distance $d^{max}$
is a linear function of confinement potential range $R$.
Figures 10 and 11 allow us to determine the parameters of the QD nanostructure, for which the exchange
energy is maximal.  As we have pointed out these parameters correspond to
the inner-outer QD nanostructure (cf. Figures 2 and 6).
For the coupled QD's separated by the potential barrier the exchange energy is considerably smaller.

\begin{figure}[htbp]
\begin{center}
\begin{turn}{-90}
\epsfig{figure=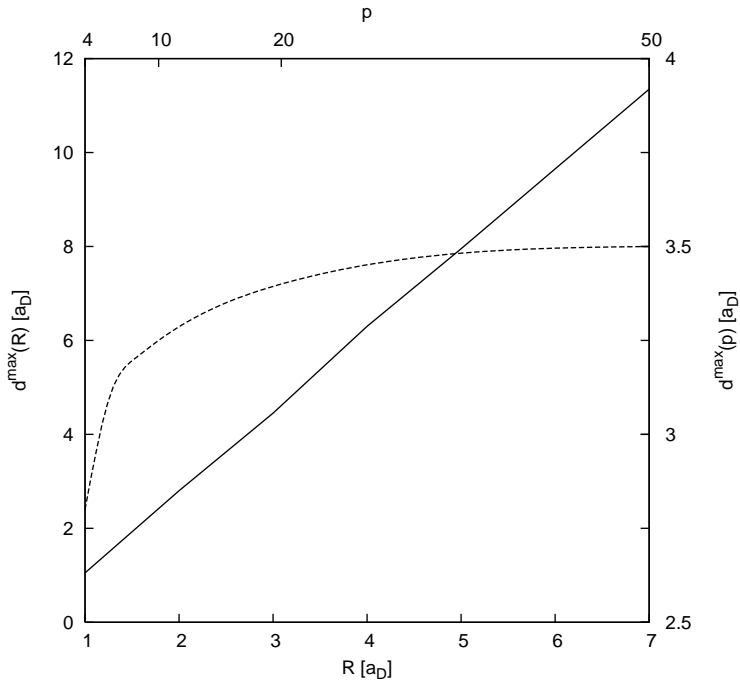,width=90mm}
\end{turn}
\end{center}
\caption{Distance $d^{max}$ between the confinement potential centres,
for which the exchange energy is maximal, as a function of
confinement potential range $R$ (solid curve for $p=4$)
and softness parameter $p$ (dashed curve for $R= 2 a_D$).}
\end{figure}

\section{Discussion}

The parametrization of the confinement potential
given in Eq.~(\ref{conf}) is sufficiently flexible
to model various types of QD's, among them the electrostatically gated QD's \cite{elz03}
and self-assembled QD's with the compositional modulation \cite{siv98}.  Using
the confinement potential (\ref{conf})
we can describe the effects of smoothness of the QD boundaries \cite{mlin07}.
For the intermediate separations $d$ between the potential-well centres
($d \simeq 3 a_D$ for $R \simeq 2 a_D$) the resulting potential profile corresponds
to the core-shell QD nanostructure with the attractive core potential well.
This profile of the confinement potential is characteristic for the compound
QD nanostructure with the inner and outer QD's, which has been recently realized
by the chemical growth of self-assembled CdTe/CdS QD's \cite{core_shell}.
The inner-outer QD nanostructure can also be realized in
the laterally coupled electrostatically gated QD's \cite{elz03}.
The corresponding profile of the confinement potential
can be obtained by applying the suitably tuned external voltages
to the two pairs of different gates.  When varying the external voltages,
we can tune the shape of the confinement potential, which in turn
leads to a desirable change of the exchange interaction.

We have shown that the exchange energy is maximal
for the compound QD's, which consists of the inner QD with the deep potential well
and outer QD with the shallow potential well.
The maximum of the exchange energy is caused by
the strong electron localization in the inner QD and the large difference
of the electron localization between the singlet and triplet states (cf. Figures 7, 8, and 9).
This result gives us a possibility of designing the nanostructures,
in which the exchange interaction is sufficiently large
to be used in quantum logic gates \cite{mosk07}.
The strong exchange interaction
is very important for performing high-fidelity quantum logic operations
with spin qubits in QD's \cite{loss98,burk99,burk00,mosk07}.
The recent study \cite{mosk07} of the exchange-interaction induced
spin swap operation shows that the swapping the electron spins
is very effective, i.e., the electron spins are fully interchanged in the possibly short time,
if the confinement potential changes from the double potential
wells separated by the barrier to the deep potential well located inside the shallow potential well.
The later potential profile corresponds to the inner-outer QD nanostructure
[cf. Figure 2(b,c)].

The exchange interaction in the coupled QD's can also be tuned by applying the external magnetic
\cite{bs04,zh06,stopa} and electric \cite{petta} fields.
The increasing magnetic field causes the decrease of the exchange interaction energy
since it lowers the energy of the triplet state and leads to the singlet-triplet degeneracy
for high magnetic fields \cite{bs04,zh06}.  The rapid changes of the external electric field
enabled Petta et al. \cite{petta} to perform a coherent manipulation of spin qubits in laterally
coupled QD's.  Szafran et al. \cite{bs04} showed that the asymmetry
of the QD's considerably increases the exchange energy.
The size effects in the exchange coupling were studied
by Zhang et al. \cite{zh07}.  The increasing size of the QD nanostructure leads to
the decrease of the exchange interaction due to the decreasing interdot tunnelling
\cite{zh07}.
Using the present results we can also determine the size effect in the exchange coupling.
In particular, Figure 10 shows that the exchange energy scales
as $\sim 1/R$ with the increasing size $R$ of the QD's.

\section{Conclusions and Summary}

In the present paper, we have studied the lowest-energy singlet and triplet two-electron states
in laterally coupled QD's and determined
the exchange interaction between the electrons.
The application of the two-centre PE function parametrization
[Eq.~(\ref{conf})]  enabled us to investigate
a large class of realistic confinement potentials with different shapes.
We focus on the dependence of the exchange
energy on the distance between the confinement potential centres
and also on the shape and range of this potential.
The dependence of the exchange energy on intercentre distance $d$
is qualitatively different for the soft ($p=2$)
and hard ($p \geq 4$) confinement potentials.
For $p=2$ the exchange energy is a monotonically decreasing
function of $d$, while for $p \geq 4$ the exchange energy
increases with $d$ for small $d$, reaches the maximum for
intermediate $d$, and next decreases to zero
for large interdot separations.
This knowledge allows us to predict the nanostructure parameters,
in particular their size and geometry,
which maximize the exchange energy.

In summary, we have found that the exchange energy is maximal for
the confinement potential, which corresponds to the compound QD nanostructure,
consisting of the inner QD with deep potential well embedded in the outer QD
with shallow potential well.  The corresponding core-shell confinement potential
can be obtained in the form of the inner-outer QD nanostructure realized
in self-assembled QD's and electrostatically gated QD's.
We have also investigated the tuning of the exchange interaction by changing
the parameters of the coupled QD nanostructure and pointed out
the importance of the present study for the quantum logic
operations with electron spins.

\section*{Acknowledgment}

We are grateful to Bart\l omiej Szafran for a technical assistance
and fruitful scientific discussions. This work has been partly
supported by the Polish Ministry for Science and High School Education.

\end{document}